\documentclass[11pt]{revtex4-1}

    \usepackage{amssymb}
    \usepackage{graphicx}
    \usepackage{amsfonts}              
    \usepackage{amssymb,amsmath}
    \usepackage{bm}
	 \usepackage[margin=1.4in]{geometry}
    \usepackage{float}
    \begin{document}
    \author{Ching Hua Lee}

\title{Holographic Topological Insulator}\maketitle
Topological Insulators are novel materials that possess robust conducting surfaces or edges despite being bulk insulators. Conceptually, they are interesting because their surface or edge states are protected by the topological properties of their bulk bandstructure. From a technological standpoint, they are also exciting due to their potential as dissipationless wire interconnects in nanoscale circuits and spintronics devices \cite{zhang2012chiral}. So far, the three types of topological insulators that have been experimentally realized are the 2D quantum spin Hall (QSH), 2D quantum anomalous Hall (QAH) and the 3D time-reversal invariant $Z_2$ topological insulator which are respectively realized, for instance, in HgTe quantum wells, Cr-doped BiSeTe thin films and BiSb.   

Parallel to these advances in topological insulators is the intense interest of holographic duality in the theoretical physics community. First proposed by Witten, Maldacena, Klebanov and others in 1998, it is also known as the Anti-de-Sitter space/Conformal Field Theory (AdS/CFT) correspondence. Its principal idea is that a quantum field on a fixed background geometry can be regarded as a "hologram" containing the same information as a "dual" gravitational system one dimension higher. As the extra emergent dimension in the dual system has the physical interpretation of scale, holographic duality enables the scaling behavior of the original system to be understood in terms of spatial dynamics in the dual spacetime. The best understood example of holographic duality is the correspondence between 4-dimensional super-Yang-Mills theory and 5-dimensional supergravity, where the strongly-coupled limit of the super-Yang-Mills theory corresponds to the classical (weakly-coupled) limit of the dual gravitational theory. In a similar flavor, holographic duality has also been fervently employed in the study of strongly-coupled condensed matter systems like quantum critical heavy Fermion systems and putative high-temperature superconductors. Indeed, by providing an alternative route to understanding the elusive behavior of strongly-coupled systems, holographic duality is a quintessential example of the symbiosis between high energy physics and condensed matter physics.

\begin{figure}
\includegraphics[scale=.47]{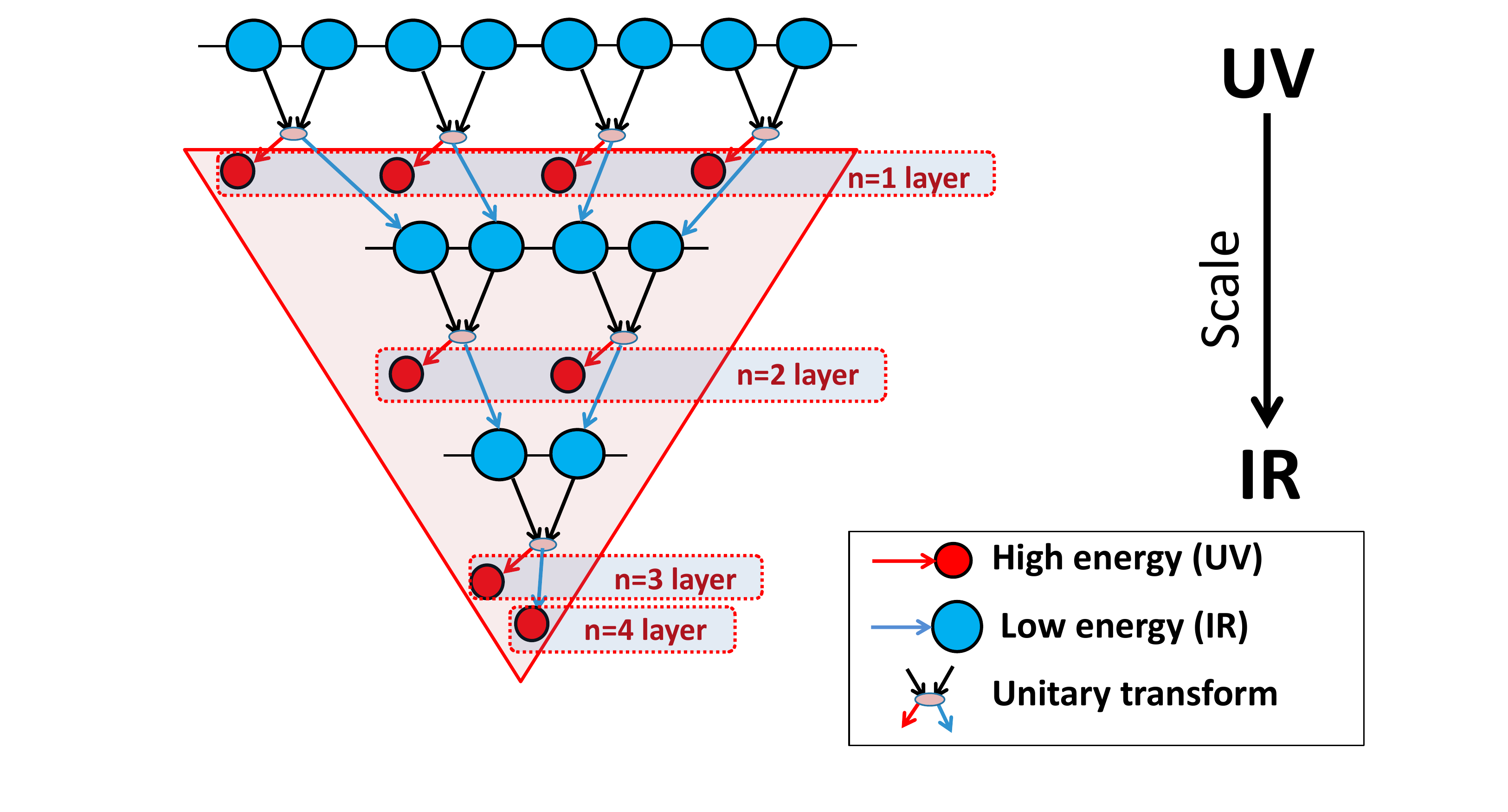}
\caption{(Color Online) A simplest illustration of the exact holographic mapping \cite{lee2016exact}. The original system consists of the $2^3=8$ sites in the top row. An unitary transform acts on the Hilbert spaces on each pair of these sites and separates them into high energy (UV, red) and low energy (IR, blue) degrees of freedom. The high energy degrees of freedom forms the first ($n=1$) layer of the dual system, while the low energy degrees of freedom form the input to the next iteration of unitary transformation. This is repeated until there is no more pair to be iterated. The result is a rearrangement of the original system into a dual system with layers of different energy scales arranged in a pyramid-like fashion.}
\label{ehmtree}
\end{figure}

Presented with illustrious developments in these two directions, it is natural to ask what holography can teach us about topology. A first step was made in \cite{qi2013}, where a systematic holographic decomposition of a lattice system was developed. This approach, known as the exact holographic mapping (EHM), allows one to write down the holographic dual of any given lattice system through repeated applications of unitary mappings (see Fig. \ref{ehmtree}). This dual possesses a Hilbert space identical to that of the original system, but is arranged in layers representing different energy scales. 

In the spirit of holography in high energy physics, one can define the classical geometry of the dual system by identifying the geodesic distance between two points on it with the upper bound of their corresponding correlation functions. As detailed in \cite{qi2013} and \cite{lee2016exact}, this upper bound is given by the mutual information between the corresponding causal cones of the two points in the original system. One finds the duals of a critical zero temperature and nonzero temperature fermionic system being Anti-de-Sitter (AdS) space and the  Ba\~{n}ados, Teitelboim  and Zanelli (BTZ) black hole respectively, in agreement with expectations from other approaches in AdS/CFT.

\begin{figure}
\includegraphics[scale=0.32]{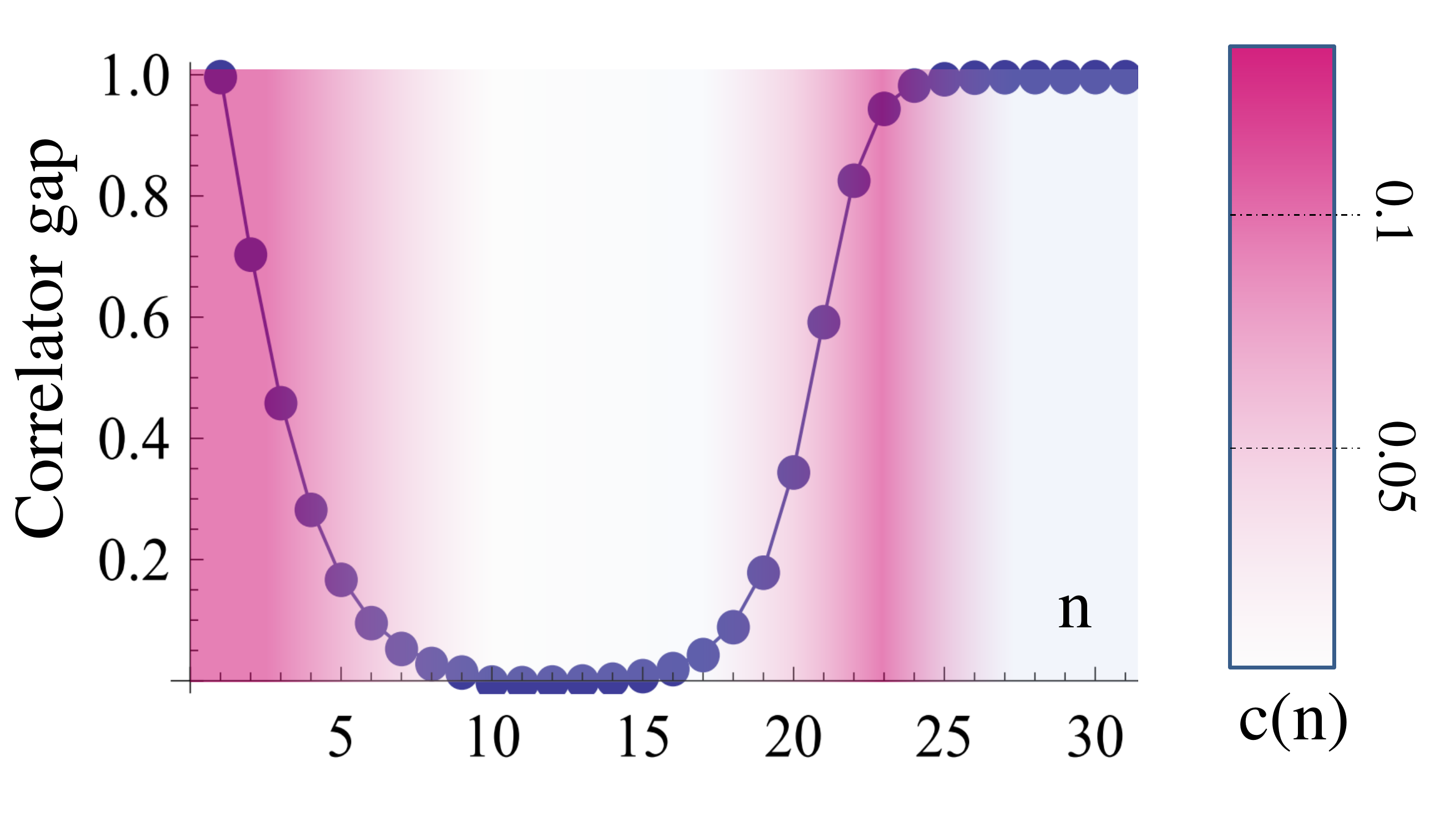}
\caption{Plot of the gap of onsite correlator in the presence of an entanglement cut at the $n^{th}$ layer. The background color density represents the Chern number density, with the two distinct left and right peaks at the scales of lattice regularization and Dirac cone gap respectively. Evidently, the correlator is gapless only if the cut is made between the two peaks, where the $Z_2$ index is purportedly nontrivial. }
\label{Holographic TI}
\end{figure}

Very interestingly, the exact holographic mapping (EHM) also provides another way of understanding the relationship between different types of topological insulators. Specifically, applying the EHM onto an almost gapless 2D quantum anomalous Hall (Chern) insulator results in a dual system containing a 3D time-reversal breaking $Z_2$ topological insulator region (Fig. \ref{Holographic TI} \cite{gulee2016}). This can be understood as follows. A 2D Chern insulator is characterized by one or more Dirac cones in momentum space, while a 3D $Z_2$ topological insulator has an odd number of Dirac cones on its spatial surfaces. Each lattice Dirac cone carries a topological flux (Chern number) of one, with half residing near its singularity and the other half arising from lattice regularization. If a Dirac cone is almost gapless, the scale of the singularity will be much smaller than that of lattice regularization. In this case, the EHM separates these two contributions into two distinct Dirac cone regions in the emergent scale direction of the dual system. The region between these Dirac cones is a bona-fide $Z_2$ topological insulator, as justified in \cite{gulee2016} by comparing entanglement spectra. At a deeper level, this reveals the helical surface states of the 3D topological insulator as direct manifestations of the parity anomaly of 2+1-D Dirac cones. Mathematically, one identifies the Berry curvature density distribution in the dual system with the gradient of the $\theta$-angle of the effective topological Axion field theory.  

Indeed, the application of the EHM on topological insulators has gone beyond the correspondence between critical theories and AdS spacetime. It forges a suggestive relationship between the two bulk-edge correspondences that has created much excitement in the physics community: Holographic duality which relates a conformal field theory on the edge with a gravitational theory in the bulk, and topological bulk-edge correspondence which relates a conformal field theory on the edge with a nontrivial topological invariant from the bulk states. 

Note: this research highlight was originally published at:\\
 \textsf{http://www.ntu.edu.sg/ias/newsletters/Documents/APPN/APPNv5n2May2016-lowres.pdf}

\bibliography{ehm}

\begin{thebibliography}{4}%
\makeatletter
\providecommand \@ifxundefined [1]{%
 \@ifx{#1\undefined}
}%
\providecommand \@ifnum [1]{%
 \ifnum #1\expandafter \@firstoftwo
 \else \expandafter \@secondoftwo
 \fi
}%
\providecommand \@ifx [1]{%
 \ifx #1\expandafter \@firstoftwo
 \else \expandafter \@secondoftwo
 \fi
}%
\providecommand \natexlab [1]{#1}%
\providecommand \enquote  [1]{``#1''}%
\providecommand \bibnamefont  [1]{#1}%
\providecommand \bibfnamefont [1]{#1}%
\providecommand \citenamefont [1]{#1}%
\providecommand \href@noop [0]{\@secondoftwo}%
\providecommand \href [0]{\begingroup \@sanitize@url \@href}%
\providecommand \@href[1]{\@@startlink{#1}\@@href}%
\providecommand \@@href[1]{\endgroup#1\@@endlink}%
\providecommand \@sanitize@url [0]{\catcode `\\12\catcode `\$12\catcode
  `\&12\catcode `\#12\catcode `\^12\catcode `\_12\catcode `\%12\relax}%
\providecommand \@@startlink[1]{}%
\providecommand \@@endlink[0]{}%
\providecommand \url  [0]{\begingroup\@sanitize@url \@url }%
\providecommand \@url [1]{\endgroup\@href {#1}{\urlprefix }}%
\providecommand \urlprefix  [0]{URL }%
\providecommand \Eprint [0]{\href }%
\providecommand \doibase [0]{http://dx.doi.org/}%
\providecommand \selectlanguage [0]{\@gobble}%
\providecommand \bibinfo  [0]{\@secondoftwo}%
\providecommand \bibfield  [0]{\@secondoftwo}%
\providecommand \translation [1]{[#1]}%
\providecommand \BibitemOpen [0]{}%
\providecommand \bibitemStop [0]{}%
\providecommand \bibitemNoStop [0]{.\EOS\space}%
\providecommand \EOS [0]{\spacefactor3000\relax}%
\providecommand \BibitemShut  [1]{\csname bibitem#1\endcsname}%
\let\auto@bib@innerbib\@empty
\bibitem [{\citenamefont {Zhang}\ and\ \citenamefont
  {Zhang}(2012)}]{zhang2012chiral}%
  \BibitemOpen
  \bibfield  {author} {\bibinfo {author} {\bibfnamefont {X.}~\bibnamefont
  {Zhang}}\ and\ \bibinfo {author} {\bibfnamefont {S.-C.}\ \bibnamefont
  {Zhang}},\ }\bibfield  {booktitle} {\emph {\bibinfo {booktitle} {SPIE
  Defense, Security, and Sensing}},\ }\href@noop {} {\ ,\ \bibinfo {pages}
  {837309} (\bibinfo {year} {2012})}\BibitemShut {NoStop}%
\bibitem [{\citenamefont {Qi}(2013)}]{qi2013}%
  \BibitemOpen
  \bibfield  {author} {\bibinfo {author} {\bibfnamefont {X.-L.}\ \bibnamefont
  {Qi}},\ }\href@noop {} {\bibfield  {journal} {\bibinfo  {journal} {arXiv
  preprint arXiv:1309.6282}\ } (\bibinfo {year} {2013})}\BibitemShut {NoStop}%
\bibitem [{\citenamefont {Lee}\ and\ \citenamefont {Qi}(2016)}]{lee2016exact}%
  \BibitemOpen
  \bibfield  {author} {\bibinfo {author} {\bibfnamefont {C.~H.}\ \bibnamefont
  {Lee}}\ and\ \bibinfo {author} {\bibfnamefont {X.-L.}\ \bibnamefont {Qi}},\
  }\href@noop {} {\bibfield  {journal} {\bibinfo  {journal} {Physical Review
  B}\ }\textbf {\bibinfo {volume} {93}},\ \bibinfo {pages} {035112} (\bibinfo
  {year} {2016})}\BibitemShut {NoStop}%
\bibitem [{\citenamefont {Gu}\ \emph {et~al.}(2016)\citenamefont {Gu},
  \citenamefont {Lee}, \citenamefont {Wen}, \citenamefont {Cho}, \citenamefont
  {Ryu},\ and\ \citenamefont {Qi}}]{gulee2016}%
  \BibitemOpen
  \bibfield  {author} {\bibinfo {author} {\bibfnamefont {Y.}~\bibnamefont
  {Gu}}, \bibinfo {author} {\bibfnamefont {C.~H.}\ \bibnamefont {Lee}},
  \bibinfo {author} {\bibfnamefont {X.}~\bibnamefont {Wen}}, \bibinfo {author}
  {\bibfnamefont {G.~Y.}\ \bibnamefont {Cho}}, \bibinfo {author} {\bibfnamefont
  {S.}~\bibnamefont {Ryu}}, \ and\ \bibinfo {author} {\bibfnamefont {X.-L.}\
  \bibnamefont {Qi}},\ }\href {\doibase 10.1103/PhysRevB.94.125107} {\bibfield
  {journal} {\bibinfo  {journal} {Phys. Rev. B}\ }\textbf {\bibinfo {volume}
  {94}},\ \bibinfo {pages} {125107} (\bibinfo {year} {2016})}\BibitemShut
  {NoStop}%
\end{thebibliography}%

\end{document}